\documentclass{IEEEtran}
\usepackage{cite}
\usepackage{amsmath,amssymb,amsfonts}
\usepackage{algorithmic}
\usepackage{graphicx}
\usepackage{textcomp}
\usepackage{color}
\usepackage{balance}
\usepackage{soul}

\begin{document}
\title{Modular Compact Modeling of Magnetic Tunnel Junction Devices}
\author{Mustafa Mert Torunbalci, Pramey Upadhyaya, Sunil A. Bhave, and Kerem Y. Camsari
\thanks{This work was supported by the National Science Foundation through the NCN-NEEDS program, under contract 1227020-EEC. 

M. M. Torunbalci, P. Upadhyaya, S. A. Bhave, and K. Y. Camsari are with 
the School of Electrical and Computer Engineering, Purdue University, West Lafayette, IN 
47907 USA, (e-mail: kcamsari@purdue.edu). }}

\maketitle

\begin{abstract}
This paper describes a robust, modular, and physics-based circuit framework to model conventional and emerging Magnetic Tunnel Junction (MTJ) devices. Magnetization 
dynamics are described by the stochastic Landau-Lifshitz-Gilbert (sLLG) equation whose results are rigorously benchmarked with a Fokker-Planck Equation (FPE) description of  magnet dynamics. We then show how sLLG is coupled to transport equations of MTJ-based devices in a unified circuit platform. Step by step, we illustrate how the physics-based MTJ model can be extended to include different spintronics phenomena, including spin-transfer-torque (STT), voltage-control of magnetic anisotropy (VCMA) and spin-orbit torque (SOT) phenomena  by experimentally benchmarked examples. To demonstrate how our approach can be used in the exploration of novel MTJ-based devices, we also present a  recently proposed MEMS resonator-driven spin-torque nano oscillator (STNO) that can reduce the phase noise of STNOs. We briefly elaborate on the use of our framework beyond conventional devices. 
\end{abstract}

\section{Introduction}
\label{sec:introduction}
\IEEEPARstart{T}{he} recent progress of spintronics in the past two decades has led to the commercial development of the STT-MRAM technology as a low power and high-speed memory device \cite{gallagher2006development, bhatti2017spintronics}. Integration of MTJs with existing CMOS devices has necessitated the development of experimentally-informed and physics-based compact  models. So far, many MTJ models with varying degrees of sophistication have been developed for circuit simulators such as SPICE and Verilog-A \cite{faber2009dynamic,guo2010spice, munira2012quasi, panagopoulos2013physics, manipatruni2014vector, jabeur2014comparison, zhang2015compact, chen2015comprehensive,  kim2015technology, camsari2015modular,  kazemi2016compact, camsari2016modular, de2017compact, modnano}. 

The objective of this paper is to present a compact, physics-based SPICE framework for MTJ-based devices that can incorporate emerging materials and phenomena in 
a modular fashion. Specifically, our approach has the following distinguishing features:
\begin{itemize}
\item Transport nodes are expressed in terms of ``spin-circuits'' that generalize ordinary charge-based networks to include spin and charge currents at each node, generalizing scalar conductances to matrices. This feature enables the modularity of our approach by directly solving for the underlying spin transport equations. 

\item Each individual module is carefully benchmarked by  a physical theory (e.g. spin diffusion equations)  and/or by an experiment with stated assumptions. 

\item The  framework can easily be extended to build new devices using existing modules, for example a GSHE (giant spin hall effect)-driven MTJ is modeled by combining the charge-driven MTJ module with a GSHE module without requiring a new model. 
\end{itemize}

We illustrate the modular circuit framework with different examples: We first validate the nanomagnet dynamics captured by the sLLG by benchmarking it against a Fokker-Planck Equation (FPE) for magnets (Section \ref{FPEvsLLG}).  Next, we briefly describe the individual transport modules (Section \ref{model}) and present an experimentally benchmarked STT-MRAM cell (Section \ref{STT-MRAM}). We show how the bias-dependent MTJ model can be extended to capture the voltage-control of magnetic anisotropy (VCMA) effect. We then switch to 3-Terminal MTJ devices (Section \ref{SOT-MRAM}) and combine the existing MTJ models with benchmarked spin-orbit modules. Lastly, we demonstrate how these modules can be combined to describe exploratory and hybrid MTJ devices using a recently proposed MEMS resonator-driven spin-torque nano oscillators (STNO) (Section \ref{MASH}). 

\begin{figure*}[!h] 
\centerline{\includegraphics[width=0.76\linewidth]{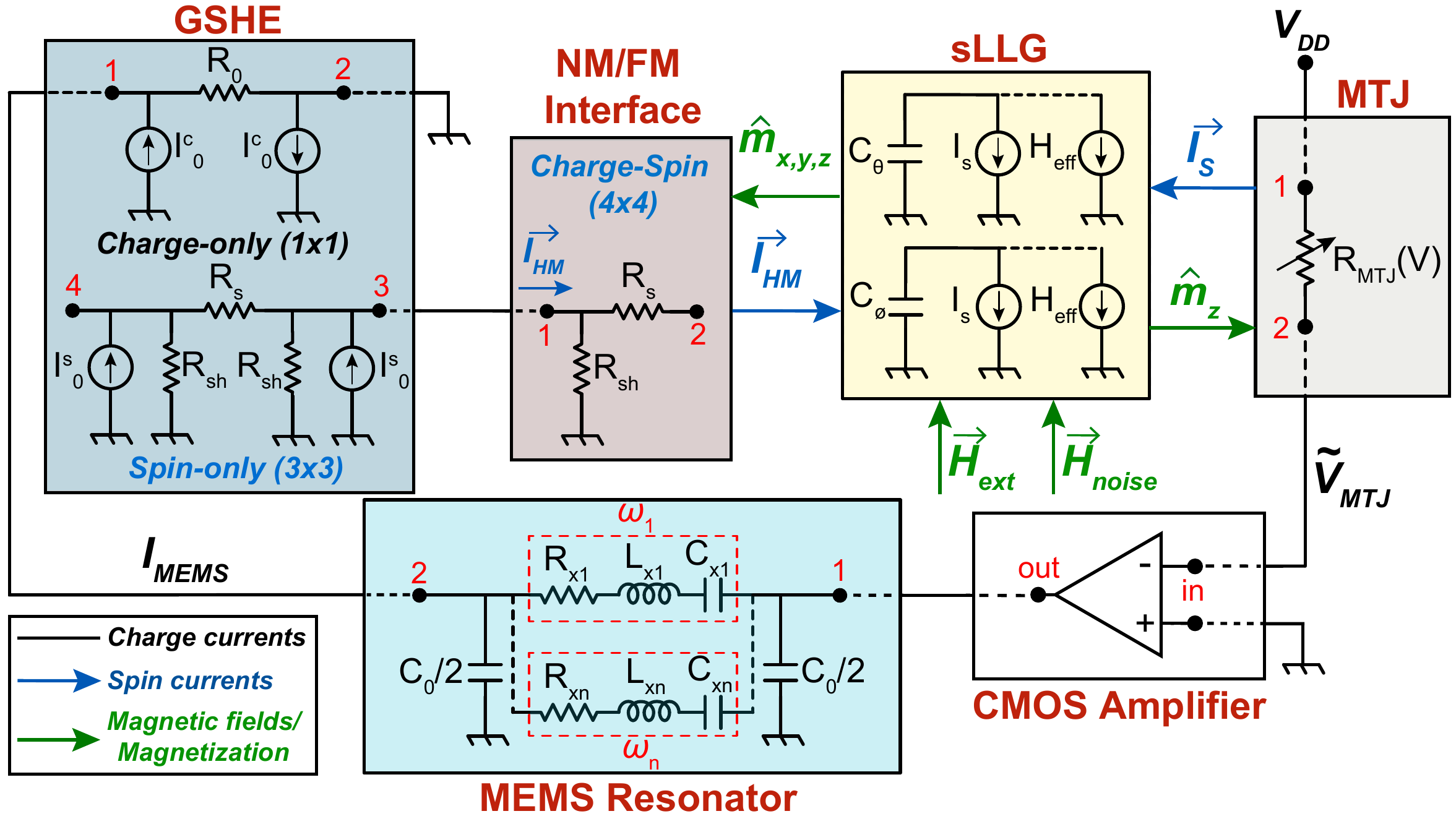}}
\caption{\textbf{Modular framework for modeling MTJ-based devices:} This figure shows explicit circuit schematics for all modules that are used in this paper and used as is to model the hybrid MEMS-STNO device discussed in Section \ref{MASH}. The circuit connections correspond to the MEMS resonator-driven MTJ-based oscillator that is described in Section \ref{MASH}. This figure also shows how each module interacts with each other. Charge currents and voltages (Black lines) are solved according to ordinary circuit theory, while spin-currents (Blue lines) and magnetization (Green lines) are provided to and from the sLLG solver respectively.  sLLG and transport models are described in Sections~\ref{FPEvsLLG} and ~\ref{model}, respectively. GSHE and NM/FM modules stand for Giant Spin Hall Effect and normal metal/ferromagnet interface. $R_s$ and $R_{sh}$ stand for series and shunt resistances.  Butterworth-Van Dyke (BVD) equivalent circuit is used to model MEMS resonators \cite{larson2000modified}.  Predictive Technology Models (PTM) are used for CMOS circuits\cite{cao2002predictive}.}
\label{Visionfig}
\end{figure*}

In Fig.~\ref{Visionfig}, we show a spin-circuit that is used to model the hybrid MEMS-STNO device discussed in Section \ref{MASH} (Fig.~\ref{MASHfig}). This example combines most of the modules discussed in this paper and we use it to show explicit circuit schematics of each module. Charge currents and voltages (Black lines) are solved according to ordinary circuit theory, while spin-currents (Blue lines) and magnetization vectors (Green lines) are provided to and from the sLLG solver respectively.  It is important to note that the time scales for magnetization dynamics are typically much shorter than typical transport times, therefore the charge circuit can be treated as a lumped model for each magnetization vector at a given time, allowing well-defined transient simulations.  Each example in this paper has been generated from the circuit models shown in Fig.~\ref{Visionfig} using the parameters defined in Table I.

\section{Validation of SPICE-based sLLG Model}  \label{FPEvsLLG}
\begin{figure}[!t]
\centerline{\includegraphics[width=0.90\linewidth]{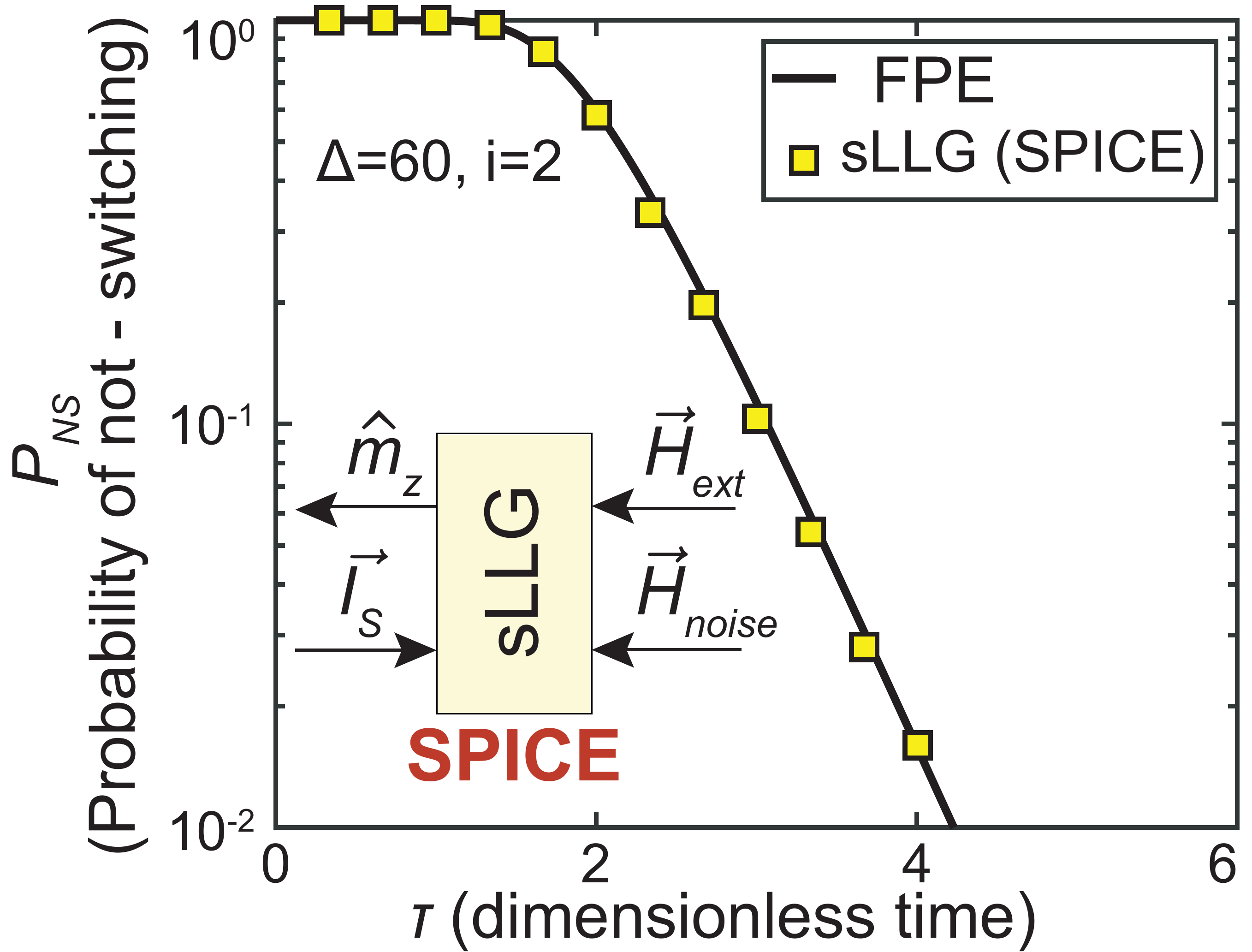}}
\caption{\textbf{FPE vs sLLG comparison}: An ensemble of magnets (N=1000) with perpendicular anisotropy with a 60 kT energy barrier is subject to a spin-current in the -z direction twice the critical spin-current. Probability of not-switching ($P_{NS}$) is obtained from a numerical solution of FPE (solid lines) and compared with a SPICE-based sLLG as explained in the text.}
\label{FPEfig}
\end{figure}

In this section, we show how the SPICE-based sLLG equation is benchmarked
by the FPE, closely following the treatment outlined in \cite{butler2012switching}.
Fig.~\ref{FPEfig} shows how the sLLG equation rigorously reproduces the FPE results. The sLLG equation is solved by the transient noise feature (.trannoise) of the HSPICE simulator, however the equations and circuits are simulator-independent and can be implemented by powerful circuit platforms such as MAPP \cite{wang2016multiphysics}. 
Our implementation uses spherical coordinates to solve the LLG equation, but the noise
field is input in Cartesian coordinates and added to any other external magnetic field
before being expressed in spherical coordinates for the LLG solver. The LLG equation
in the monodomain approximation reads:
\begin{eqnarray}
 (1+\alpha^2)\frac{d\hat{m}_i}{dt}=-|\gamma|d\hat{m}_i\times \vec{H}_{eff} \nonumber \\ -\alpha|\gamma|(\hat{m}_i \times \hat{m}_i \times \vec{H}_{eff})+ \nonumber \\ 
\frac{1}{qN_s}(\hat{m}_i \times\vec{I}_{s}\times\hat{m}_i  )+\frac{\alpha}{qN_s}(\hat{m}_i \times\vec{I}_{s})
 \label{eq:llg}
\end {eqnarray}
where $\hat{m}_i$ is the unit vector along the magnetization, $\gamma$ is the gyromagnetic ratio, $\alpha$ is the damping constant,  and $\vec{I}_{s}$ is the total spin current.  $N_s$ is the total number of spins given by $N_s$=$M_s$$\rm V_{free}$/$\mu_B$ where $M_s$ is the saturation magnetization, $\rm V_{free}$ is the volume of the free layer, $\mu_B$ is the Bohr magneton, and  $\vec{H}_{eff}$ is the effective magnetic field including the uniaxial, shape anisotropy and magnetic thermal noise terms. The magnetic thermal noise ($\vec{H}_{n}$) enters as an additional magnetic field in three dimensions with the following mean and variance, uncorrelated in all three directions:
\begin{equation}
\big \langle {H}^{x,y,z}_{n} \big \rangle=0 \quad \textrm{and} \quad   \big \langle ({H}^{x,y,z}_{n})^2 \big \rangle= \displaystyle\frac{2\alpha kT}{\gamma M_s V}
\end {equation}
 \label{eq:noise}
where $k$ is the Boltzmann constant and $T$ is the temperature.  The model becomes more accurate as devices are scaled down, maintaining the single domain approximation.

As shown in Ref. \cite{butler2012switching}, FPE reduces to a simple boundary value problem for magnets with perpendicular
anisotropy, as long as all the external fields and currents are confined to the direction of the easy-axis ($\pm z$ direction). FPE describes the dynamics of a \emph{probability density} while the sLLG tracks the dynamics of a single nanomagnet and requires ensemble averaging to be compared
with FPE. 1D FPE reads:
\begin{eqnarray}
&&\frac{\partial \rho(m_z, \tau)}{\partial \tau} = \nonumber \\
 && \frac{\partial}{\partial m_z}\left[(i-h-m_z)(1-m_z^2)\rho + \frac{1-m_z^2}{2\Delta}\frac{\partial \rho}{\partial m_z}\right]
 \label{eq:fpe}
\end{eqnarray} 
where $\tau$ is a normalized time in terms of the damping coefficient $\alpha$, the uniaxial anisotropy constant $H_k$
and the gyromagnetic ratio $\gamma$ such that $t = \tau (1+\alpha^2)/(\alpha \gamma H_k)$ and $t$ denotes real time. $\Delta$ represents the normalized
energy barrier of the magnet $(\Delta) kT$ representing the total energy barrier. The input spin current, $i$ and the external
magnetic field, $h$ are both in the $\pm z$ direction, and they are normalized by $I_{sc}$ and $H_k$ respectively, where 
$I_{sc}$ is given by $4 q/ \hbar \alpha \Delta (kT)$. When the results of FPE and sLLG are compared
for a spin-current input $i$, the corresponding current $i \times I_{sc}$ must be used for sLLG. 

Since Eq.~\ref{eq:fpe} is a 1D Boundary Value Problem, we can solve it using the bvp4c function of MATLAB
subject to the boundary condition $\partial \rho(m_z,\tau)/\partial m_z = 0$ at $m_z=\pm 1$, and subject
to the initial condition: 
\begin{equation}\rho(m_z,\tau=0)= \frac{1}{Z} \exp \left[-\Delta (1-m_z^2)\right] \Theta(m_z)
\end{equation} 
where $Z$ is a normalization constant ensuring $\int  dm_z \ \rho(m_z)=1$ in the range $(-1,+1)$ and $\Theta(m_z)$ is the Heaviside function so that the  symmetric magnetization
probability is broken to start from a given distribution, in this case chosen to be close to $m_z=1$. Note that the full FPE equation requires a more general PDE solver (2D), for example, in the case of an in-plane magnet that contains fields in directions other than the easy-axis \cite{xie2017fokker}.

The procedure of comparing FPE with sLLG is as follows:
\begin{itemize}
\item Prepare 1000 identical samples that start at $m_z= +1$ at $t=0$ for the sLLG simulation. 
\item Wait for 5 ns for identical samples to thermalize and approximately form a Boltzmann distributed initial ensemble.
\item Apply spin-current in the -z direction ($I_s = i \times I_{sc}, i=2$)  at 5 ns. 
\item Measure the \textit{last} $m_z$ value at each time point for  each ensemble.
\item Obtain $P_{NS}(\tau)$ by counting the number of samples with $m_z$ values that are greater than 0 (Not-switched) and normalizing
to sample size, N=1000 (Each yellow square in Fig.~\ref{FPEfig}). 
\item Solve FPE numerically as a function of $\tau$ to obtain $\rho(m_z,\tau)$.
\item Obtain $P_{NS}(\tau)$ from the integral: 
\begin{equation}
\displaystyle\int_{m_z=0}^{m_z=+1} dm_z  \ \rho(m_z,\tau)
\end{equation}
\item Compare $P_{NS}$-FPE with $P_{NS}$-LLG (Fig.~\ref{FPEfig}).
\end{itemize}

\vspace{5pt} Note how in the second step, the requirement of a Boltzmann distributed initial ensemble is obtained naturally by running the sLLG equation for a few nanoseconds, 
so that each sample receives a random initial angle sampled from the correct equilibrium distribution, in accordance with the initial condition used for the FPE. 

This precise agreement between the FPE method and the sLLG (Fig.~\ref{FPEfig}) despite the elaborate comparison of the two completely different methods establishes the validity of our HSPICE-based sLLG model. This result is consistent with careful studies that have investigated the numerical solution of the sLLG equation \cite{garcia1998langevin, roy2016compact,ament2016solving}. 
\section{Description of Transport Models} \label{model}
In this section, we describe the transport models that are solved self-consistently with the sLLG model that takes spin-currents, either due to spin-transfer torques (STT) or spin-orbit-torques (SOT), as an input.  

\textit{Giant Spin Hall Effect:} To model the GSHE induced spin-orbit currents, we use a spin-circuit description whose results are shown to be equivalent to the spin-diffusion equations \cite{hong2016spin}. Under short-circuit (spin-sink) conditions, this model produces a spin-current that is proportional to a charge current $I_c$ 
\begin{equation}
\beta=\frac{I_{HM}}{I_{c}}=\theta\frac{l_{HM}}{t_{HM}}\bigg(1-\sec\bigg[\frac{t_{HM}}{ \lambda_{sf}}\bigg]\bigg)
 \label{eq:she}
\end {equation}
where $\theta$ is the spin Hall angle, $l_{HM}$ and $t_{HM}$ are the length and thickness, and $\lambda_{sf}$ is the spin-flip length of the heavy metal. The GSHE module is implemented as two coupled circuits, one describing the charge transport (in the longitudinal direction) and the other describing the spin transport (in the transverse direction) in terms of 3-dimensional spin currents and voltages \cite{hong2016spin}. 

\textit{Normal Metal (NM)/Ferromagnet (FM) interface}: This module represents the interface between a normal metal and a ferromagnet in terms of the microscopic ``mixing conductances'' that describe the transmission and reflection of spin-currents incident to an FM from a metal \cite{brataas2006non, srinivasan2014modeling}. For sufficiently large mixing conductances the incident spin-current is transferred to the ferromagnet with 100\% efficiency (spin-sink). Combined with the GSHE module, the interface circuit can reduce the spin-current efficiency to less than 100\% and can capture subtle physical effects such as spin-hall magnetoresistance \cite{hong2016spin}. The details of the interface between a heavy metal and the ferromagnet is believed to play an important role as recent studies suggest \cite{shi2018fast}, and the NM/FM interface circuit can be modified to include these phenomena in the future.

\textit{MTJ:} The bias dependence of the MTJ is captured by assuming voltage dependent interface polarizations $P(V)$. This observation is supported by a microscopic Non-Equilibrium Green's function based quantum transport model \cite{datta2012voltage, datta2017quantitative}. This physically motivated assumption may not precisely fit the bias-dependence of MTJs for any given device, however, we believe that it can be useful in modeling more complicated junctions involving multiple interfaces\cite{double_int_2013}.

\begin{figure}[t!]
\centerline{\includegraphics[width=0.95\columnwidth]{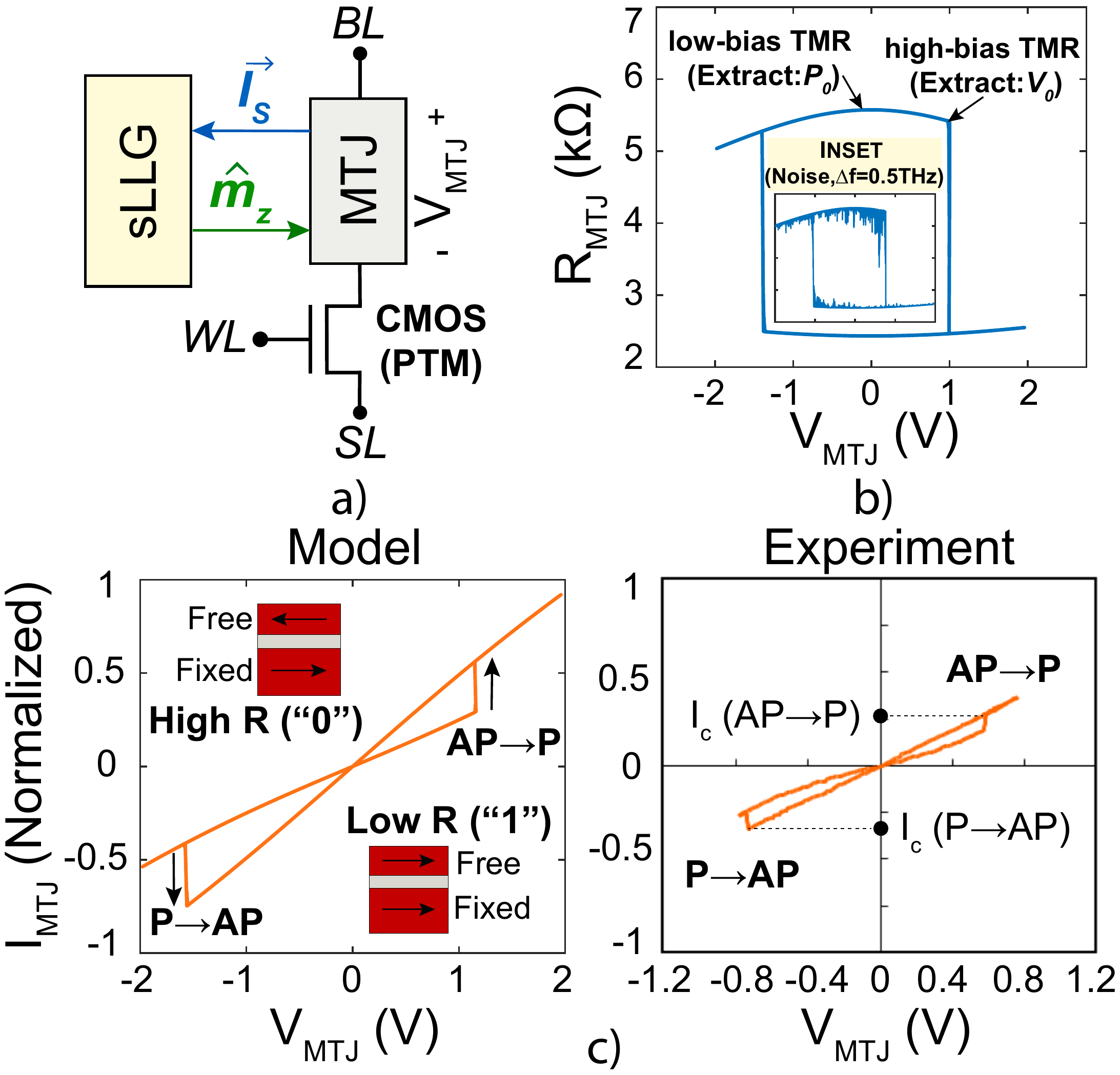}}
\caption{\textbf{STT-MRAM \cite{lin200945nm}:} (a) Spin-circuit modeling of a STT-MRAM cell, consisting of a 2-Terminal MTJ in series with an access transistor. (b) Bias-dependent TMR is used to extract the parameters $P_0$ and $V_0$ of Eq.~\ref{eq:pol}. Inset shows a stochastic simulation for one sample.  (c) Experimental benchmarking of the STT-MRAM.  In the simulation, MTJ voltage is swept from negative to positive while I$-$V characteristics are being monitored. Therefore, the circuit produces the I$-$V characteristics in one long transient simulation solving sLLG and transport equations self-consistently. Experimental figure is reprinted with permission from \cite{lin200945nm}.}
\label{STT-MRAMfig}
\end{figure}

In this picture, the MTJ conductance is expressed as:
\begin{equation}
G_{MTJ}(V)=G_0 \left[1+P_1(V)P_2(-V) \mathrm{cos(\theta)}\right]
\end{equation}
where $G_{\rm 0}$ is average MTJ conductance ($G_{\rm P}$+$G_{\rm AP}$)/2, $V$ is the voltage drop across the junction and $P_1(V)=P_2(-V)$ are the voltage dependent interface polarizations for the free and fixed layers respectively, assuming a symmetric junction. $\theta$ describes the angle between the fixed layer and the free layer. We postulate the voltage dependent polarizations as:
 \begin{equation}
  P_{1,2}(V)=\frac{1}{1+P_0 \exp(-V/V_0)} \label{eq:pol}
   \end{equation} 
where the parameter $P_0$ is determined by the low-bias Tunneling Magnetoresistance (TMR) and $V_0$ is determined by the high-bias features of TMR (Fig.~\ref{STT-MRAMfig}). $V_0$ is assumed to be the same for fixed and free layer interfaces assuming symmetric junctions throughout this paper. This functional form of the bias dependent interface polarization could  explain the observed switching asymmetry between V(P$\rightarrow$AP) and V(AP$\rightarrow$P) \cite{datta2012voltage} and the bias dependence of in-plane and out-of-plane torques. In our model, the spin-current (in-plane) incident to the free layer is  given by: $\vec{I}_S = P_2(V)(G_0) \hat M$ where $P_2$ is the interface polarization of the fixed layer, $G_0$ is the average MTJ conductance and $\hat M$ is the magnetization direction of the fixed layer as discussed in \cite{datta2012voltage}. Physically, the spin-current incident to the free layer is proportional to the polarization of the fixed layer, capturing the angular dependence of the in-plane spin-currents \cite{camsari2014physics}. The out-of-plane spin-current can similarly be included but we assume it to be zero in the examples shown in this paper. 

\section{STT-MRAM} \label{STT-MRAM}
A simple STT-MRAM cell consists of a 2-Terminal MTJ in series with an access transistor (1T/1MTJ) where the data is stored in the free layer (FL) magnetization.  Unlike traditional MRAM that uses magnetic fields, the FL magnetization is switched by passing a charge current through the fixed layer that gets spin polarized for writing information.  This can be done by applying proper voltages to the bit line (BL) and the source line (SL) while keeping the access transistor ON with the appropriate word line WL voltage (Fig.~\ref{STT-MRAMfig}).  

In this section, we present an experimental benchmarking of an STT-MRAM cell \cite{lin200945nm} that consists of 2-Terminal MTJ with an access transistor embedded into a 45 nm CMOS technology. We use 14-nm High Performance-FinFET models using Predictive Technology Models \cite{cao2002predictive} with no significant difference in the final results.  

For all MTJs described in this paper, our methodology to reproduce the MTJ characteristics is as follows:
 \begin{itemize}
 \item  Use low-bias TMR to obtain the parameter $P_0$ from Julliere's formula  TMR$= \displaystyle{2 P^2(V=0)}/{[1-P^2(V=0)]}$.
 \item  Use high-bias TMR to obtain the parameter $V_0$.
 \item  Predict switching characteristics \textit{only using} $P_0$, $V_0$ and magnet parameters with no additional fitting parameters.
 \end{itemize}
  
Fig.~\ref{STT-MRAMfig} (a-b) presents the spin-circuit model and experimental benchmarking of the STT-MRAM cell in \cite{lin200945nm}.  Due to missing material information in this experiment, we assume CoFeB parameters for free and fixed layers and use typical values for $ M_s$, $H_d$, and $\alpha$ as shown in Table~I, obtaining reasonable agreement with the experiment.  Our model captures the bias-dependence of TMR  as well as the in-plane torque asymmetry by using the voltage dependent interface polarizations $P_1(V)$ and $P_2(-V)$ resulting in a higher write current for P to AP switching  compared to AP to P switching \cite{datta2012voltage}. 
   
\section{Voltage Controlled Magnetic Anisotropy}\label{vcma}
\begin{figure}[t!]
\centerline{\includegraphics[width=0.95\columnwidth]{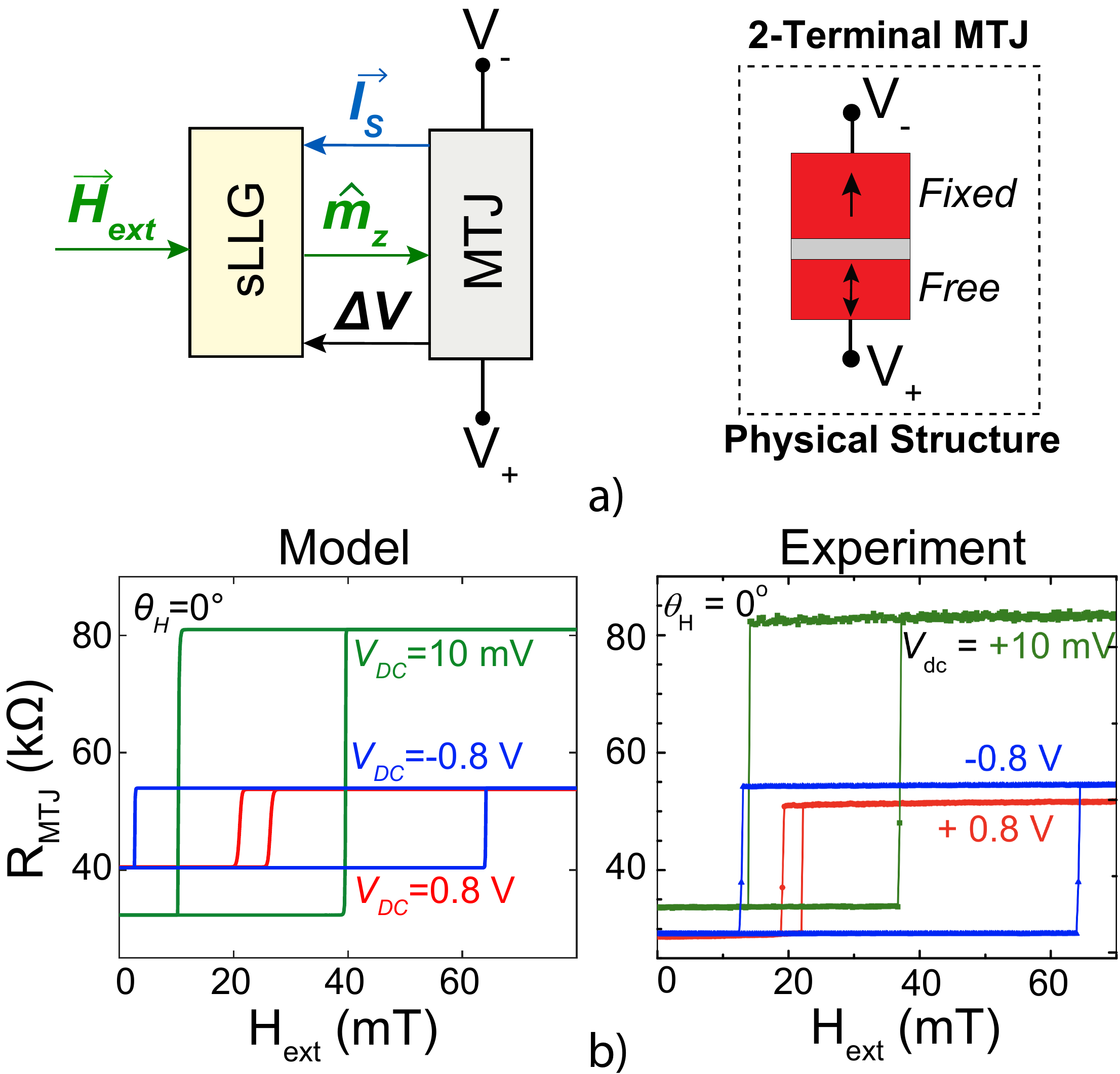}} 
\caption{\textbf{VCMA \cite{kanai2012electric}:} a) Spin-circuit model for the VCMA effect. LLG solver receives the voltage across the MTJ as an additional input to modulate the magnetic anisotropy. This change in anisotropy is modeled as a magnetic field that is dependent on instantaneous magnetization $ \Delta H_k m_z $. b) Experimental benchmarking of the VCMA effect at different voltages. A 25 mT constant dipolar shift is added. The shifts in the central dipolar fields at different voltages are due to the spin-current flowing in the MTJ.  Experimental figure is reprinted with permission from \cite{kanai2012electric}.}
\label{VCMAfig}
\end{figure}

In recent years, enormous progress has been made in the the voltage control of magnetism. A notable example is the voltage controlled magnetic anisotropy (VCMA) effect where the  electric field across a 2-Terminal MTJ, in particular with PMA magnets, induces a change in the magnetic anisotropy (See, for example \cite{kanai2012electric, amiri2012voltage}). Fig.~\ref{VCMAfig} (a-b) shows how the bias-dependent MTJ model discussed in Section~\ref{STT-MRAM} can be extended to include the VCMA effect. 

The coercivity of the PMA free layer under low-bias conditions is likely to depend on the applied pulse width and thermal activation. For simplicity, we assume this coercivity is equal to
the uniaxial anisotropy ($H_k$) in a monodomain approximation while noting that this assumption could underestimate the extracted VCMA coefficient. We assume a linear modification of the free layer anisotropy with respect to voltage: 
 \begin{equation}
  H_{k}(V)=H_{k0}+\frac{V}{t_{ox}}\left(\frac{\eta}{t_{fm} M_s}\right)
   \end{equation} 
   where $t_{fm}, t_{ox}$ are the oxide and free layer thickness and $\eta$ is the VCMA coefficient (in units $\mu \rm J / m^2$ per V/nm) and $V$ is applied voltage. Combined with the bias-dependent MTJ,  our VCMA model seems to capture three distinct effects as seen in Fig.~\ref{VCMAfig}: a) the $H_k$ modulation of the free layer, b) the bias dependence of the MTJ where $\pm 0.8 $ V decreases the TMR, and c) the shift of the hysteresis loop that arises due to a spin-current flowing from the fixed layer to the free layer in opposite directions for opposite voltage polarity. The asymmetric shift for $\pm 0.8$ V seems consistent with the asymmetry in the magnitude of the spin-current captured by the voltage dependent polarizations. Note that only the first effect is added explicitly, the rest follows from the MTJ model discussed in Section \ref{STT-MRAM}.
   \begin{figure}[t!]
\centerline{\includegraphics[width=0.925\columnwidth]{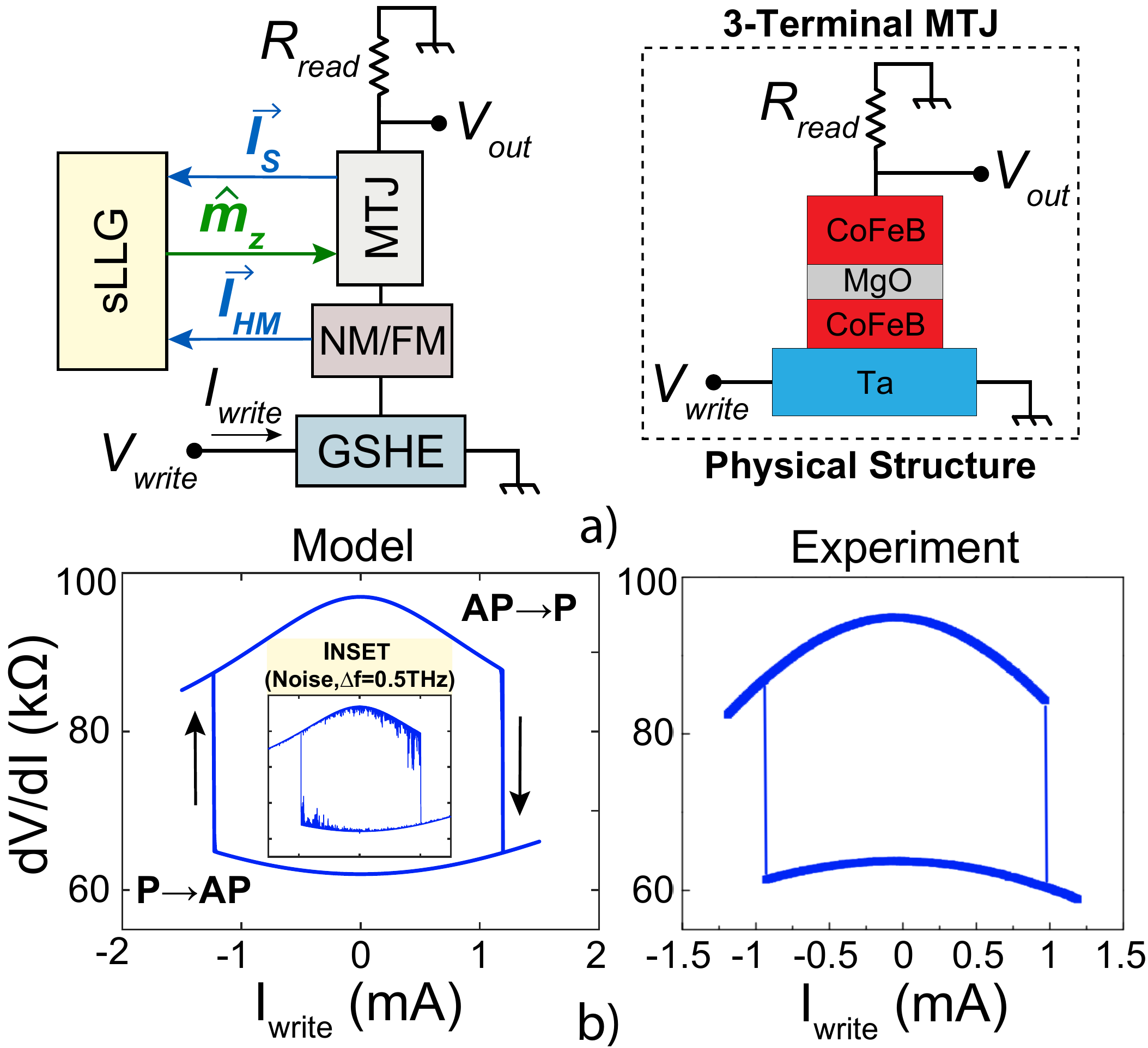}} 
\caption{\textbf{GSHE-driven MTJ \cite{liu2012spin}:} (a) Spin-circuit model and physical implementation of a GSHE-driven MTJ.  (b)  Experimental benchmarking of the GSHE-driven MTJ.  In the simulation, current through the HM is slowly swept from negative to positive while MTJ resistance is being monitored, as in previous examples. Parameters $P_0$ and $V_0$ are obtained from bias-dependence of TMR (Inset shows a stochastic simulation for one sample). Experimental figure is reprinted with permission from \cite{liu2012spin}.} 
\label{SOT-MRAMfig}
\end{figure}
\section{GSHE-driven MTJ} \label{SOT-MRAM}
An SOT-MRAM cell consists of a 3-Terminal MTJ and access transistors where spin current is generated by passing a charge current through a heavy metal for writing information.  Thus, SOT-MRAM achieves read and write operations using separate paths, improving a feature of STT-MRAMs that use same paths for read and write.  In this section, we experimentally benchmark a GSHE-driven MTJ (without the transistor periphery) that consists of an in-plane  CoFeB for fixed and free layers and Tantalum for the heavy metal (HM).   Fig.~\ref{SOT-MRAMfig} (a) shows spin-circuit modeling and physical implementation of the GSHE-driven MTJ in \cite{liu2012spin}.  We follow the same methodology to reproduce the MTJ characteristics explained in the section \ref{STT-MRAM} using  identical material and device parameters as in the experiment. Fig.~\ref{SOT-MRAMfig} (b) compares our results with the experiment that seem to be in good quantitative agreement without any additional fitting parameters. We note, however, the switching currents in the experiment could also be influenced by thermal activation mechanisms and multi-domain features that are not included in our model. The symmetric switching currents as well as the roll-off due to high bias MTJ properties are captured,  although the pronounced (and opposite) roll-off in the parallel TMR branch suggests mechanisms beyond what is in our model, possibly related to multi-domain features of the nanomagnets. 

\section{MEMS Resonator Driven STNO} \label{MASH}
\begin{figure}[t!]
\centerline{\includegraphics[width=\columnwidth]{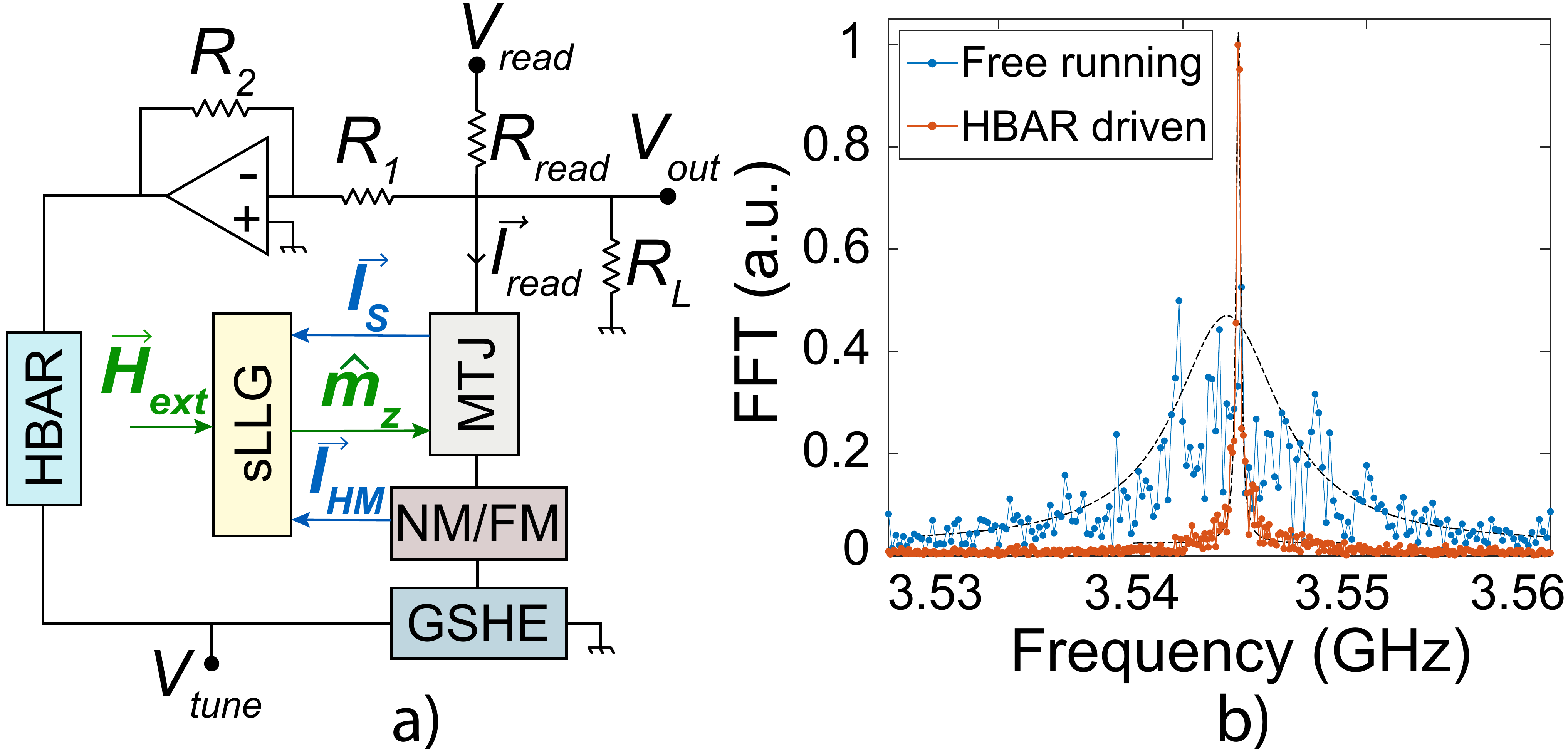}} 
\caption{\textbf{HBAR-driven STNO with current feedback: } a) Spin-circuit modeling of a MEMS-resonator driven STNO. The frequency spectrum is obtained by taking the FFT of the time domain signal.  (b) FFT spectra of free running STNO and HBAR driven STNO. A Lorentzian function is used to fit the FFT data to extract the linewidth of the oscillators. HBAR driven STNO shows $\sim$15X enhancement in the quality factor.}
\label{MASHfig}
\end{figure}
In this section, our purpose is to show how the circuit framework that we present can be used to evaluate exploratory devices, using a recently proposed hybrid MTJ-based device as an example. It is well-established that the free layer of an MTJ can be driven to function as a nano oscillator by STT and SOT mechanisms \cite{chen2016spin}. However, these spin-torque-nano oscillators (STNO) suffer from very poor phase noise characteristics\cite{quinsat2010amplitude}, limiting their potential as nanoscale, tunable oscillators. 

A recent proposal suggested the coupling of GSHE-driven STNOs with very high quality MEMS-resonators as a CMOS-compatible solution to improve the phase noise characteristics of STNOs \cite{torunbalci2018magneto}.  This concept is tested by combining a high overtone bulk acoustic wave resonator (HBAR) and 3-Terminal MTJ-based STNO. The 2-port HBAR is described by the Butterworth-Van Dyke (BVD) equivalent circuit \cite{larson2000modified}, coupled to the GSHE-driven MTJ with current feedback.  The full spin-circuit corresponding to this example is shown in Fig.~{\ref{Visionfig}}. We use similar parameters for the GSHE-driven MTJ that is described in Section \ref{SOT-MRAM}.  Fig.~\ref{MASHfig} presents the hybrid model and compares FFT spectra of a free running STNO and an HBAR-driven STNO, where HBAR-driven STNO exhibits a significant enhancement in the oscillator linewidth. Furthermore, the original tunability of the STNO is maintained by locking it to the nearest high Q peak of the HBAR. This example demonstrates how our modular framework can be combined with established circuit models to propose and evaluate hybrid devices involving MTJs.

\begin{table}[!ht] 

\caption{Parameters used for all simulations}
\label{SOT-table}
\centering 
\resizebox{0.95\columnwidth}{!}{
\begin{tabular} {|l|l|l|}
\hline
\multicolumn{2}{|c|}{STT-MRAM \cite{lin200945nm}}\\
\hline
$\rm V_{free}$ & 100 nm $\times$ 40 nm $\times$ 1.5 nm \\

$M_{s}$,  $\alpha$, $H_{k}$, $H_{d}$ & 1100 emu/cc, 0.01, 150 Oe, 1.3 T \\

TMR,  $P_{0}$, $V_{0}$ & 110\%, 0.6887, 1.81\\

\rm CMOS models, Size & 14 nm HP-FinFET, nfin=50 \\

$\rm t_{sim}$, $\rm t_{step}$ (HSPICE, noise:on)& 5 $\mu$s, 2 ps  \\
\hline

\multicolumn{2}{|c|}{VCMA \cite{kanai2012electric}}\\
\hline
$\rm V_{free}$ & 70 nm $\times$ 70 nm $\times$ 1.8 nm \\

$M_{s}$,  $\alpha$, $H_{k}=H_{ks}-H_d$ & 1100 emu/cc, 0.075, 120 Oe \\

TMR,  $P_{0}$, $V_{0}$ & 150\%, 0.5253, 0.33\\

$t_{ox}$,  $\eta$ & 1.4 nm, 6.5 $\mu \rm J / m^2$/V/nm \\

$\rm t_{sim}$, $\rm t_{step}$ (HSPICE, noise:off)& 25 $\mu$s, 5 ps  \\
\hline

\multicolumn{2}{|c|}{GSHE-driven MTJ \cite{liu2012spin}} \\
\hline
$\rm V_{free}$ & 350 nm $\times$ 100 nm $\times$ 1.6 nm \\

$M_{s}$, $ \alpha$, $H_{k}$, $H_{d}$ & 1100 emu/cc, 0.021, 40 Oe, 0.76 T \\

TMR, $P_{0}$, $V_{0}$, $\rm Re[G_{mix}]$ & 50\%, 1.13, 0.083, $\rm 10^{15}$ $\rm S/m^2$\\

$\theta _{HM}$, $\lambda_{HM}$, $\rm \rho_{HM}$ (HM=Ta)\ & 0.12, 1.5 nm, 190 $\rm \mu$$\Omega -cm$ \\

$l_{HM}$, $w_{HM}$, $t_{HM}$ & 1 $\mu$m, 5  $\mu$m, 6.2 nm\\
  
$\rm t_{sim}$, $\rm t_{step}$, (HSPICE, noise:on)& 10 $\rm \mu$s, 2 ps  \\

\hline 
\end{tabular}}
\end{table}

\section{Beyond conventional devices}
The modular framework we present in this paper can be used beyond compact modeling MTJ devices that are envisioned in the context of memory applications \cite{bhatti2017spintronics}. For example, materials such as Topological Insulators, exhibiting spin-momentum locking (SML) have been expressed in terms of general spin-circuit models recently \cite{sayed2016multi,sayed2017transmission}. In conjunction with the nanomagnet modules discussed in this paper, such SML models have been used in an exploratory mode to propose new memory devices \cite{sayed2017proposal} and to model novel devices exhibiting three resistance states \cite{sayed2016multi} that have subsequently received experimental confirmation \cite{lee2018multi}.  In addition, the benchmarked stochastic LLG solver coupled with MTJ modules has been used to model a special type of superparamagnetic MTJs \cite{vodenicarevic2017low} functioning as three-terminal tunable random number generators \cite{camsari2017implementing}, or probabilistic bits (p-bits), and p-circuits built out of such p-bits have been shown to be useful for a wide range of applications such as Ising computing \cite{sutton2017intrinsic}, Bayesian inference \cite{behin2016building} and invertible logic\cite{camsari2017stochastic}. 

\section{Conclusion}
We present a modular, extensible and physics-based circuit framework for MTJ devices that can capture a wide range of spintronic phenomena. We start by a rigorous validation of the stochastic LLG model and then demonstrate our approach with a step by step extension of our framework. The \textit{same} basic models are used to capture bias-dependence of MTJs to model STT-MRAM, GSHE driven-MTJs, VCMA-based MTJs as well as hybrid circuits such as MEMS resonator-driven STNOs.  We believe that the modular framework presented here can be an important compact modeling toolbox in the exploration of new MTJ devices as the fields of spintronics and magnetism progress with new materials and phenomena.

\bibliographystyle{IEEEtran} 

\begin{thebibliography}{10}
\providecommand{\url}[1]{#1}
\csname url@samestyle\endcsname
\providecommand{\newblock}{\relax}
\providecommand{\bibinfo}[2]{#2}
\providecommand{\BIBentrySTDinterwordspacing}{\spaceskip=0pt\relax}
\providecommand{\BIBentryALTinterwordstretchfactor}{4}
\providecommand{\BIBentryALTinterwordspacing}{\spaceskip=\fontdimen2\font plus
\BIBentryALTinterwordstretchfactor\fontdimen3\font minus
  \fontdimen4\font\relax}
\providecommand{\BIBforeignlanguage}[2]{{%
\expandafter\ifx\csname l@#1\endcsname\relax
\typeout{** WARNING: IEEEtran.bst: No hyphenation pattern has been}%
\typeout{** loaded for the language `#1'. Using the pattern for}%
\typeout{** the default language instead.}%
\else
\language=\csname l@#1\endcsname
\fi
#2}}
\providecommand{\BIBdecl}{\relax}
\BIBdecl

\bibitem{gallagher2006development}
\BIBentryALTinterwordspacing
W.~J. Gallagher and S.~S. Parkin, ``Development of the magnetic tunnel junction
  mram at ibm: From first junctions to a 16-mb mram demonstrator chip,''
  \emph{IBM Journal of Research and Development}, vol.~50, no.~1, pp. 5--23,
  2006. [Online]. Available: \url{https://doi.org/10.1147/rd.501.0005}
\BIBentrySTDinterwordspacing

\bibitem{bhatti2017spintronics}
\BIBentryALTinterwordspacing
S.~Bhatti, R.~Sbiaa, A.~Hirohata, H.~Ohno, S.~Fukami, and S.~Piramanayagam,
  ``Spintronics based random access memory: a review,'' \emph{Materials Today},
  2017. [Online]. Available: \url{https://doi.org/10.1016/j.mattod.2017.07.007}
\BIBentrySTDinterwordspacing

\bibitem{faber2009dynamic}
\BIBentryALTinterwordspacing
L.-B. Faber, W.~Zhao, J.-O. Klein, T.~Devolder, and C.~Chappert, ``Dynamic
  compact model of spin-transfer torque based magnetic tunnel junction (mtj),''
  in \emph{Design \& Technology of Integrated Systems in Nanoscal Era, 2009.
  DTIS'09. 4th International Conference on}.\hskip 1em plus 0.5em minus
  0.4em\relax IEEE, 2009, pp. 130--135. [Online]. Available:
  \url{https://doi.org/10.1109/DTIS.2009.4938040}
\BIBentrySTDinterwordspacing

\bibitem{guo2010spice}
\BIBentryALTinterwordspacing
W.~Guo, G.~Prenat, V.~Javerliac, M.~El~Baraji, N.~De~Mestier, C.~Baraduc, and
  B.~Dieny, ``Spice modelling of magnetic tunnel junctions written by
  spin-transfer torque,'' \emph{Journal of Physics D: Applied Physics},
  vol.~43, no.~21, p. 215001, 2010. [Online]. Available:
  \url{https://doi.org/10.1088/0022-3727/43/21/215001}
\BIBentrySTDinterwordspacing

\bibitem{munira2012quasi}
\BIBentryALTinterwordspacing
K.~Munira, W.~H. Butler, and A.~W. Ghosh, ``A quasi-analytical model for
  energy-delay-reliability tradeoff studies during write operations in a
  perpendicular stt-ram cell,'' \emph{IEEE Transactions on Electron Devices},
  vol.~59, no.~8, pp. 2221--2226, 2012. [Online]. Available:
  \url{https://doi.org/10.1109/TED.2012.2198825}
\BIBentrySTDinterwordspacing

\bibitem{panagopoulos2013physics}
\BIBentryALTinterwordspacing
G.~D. Panagopoulos, C.~Augustine, and K.~Roy, ``Physics-based spice-compatible
  compact model for simulating hybrid mtj/cmos circuits,'' \emph{IEEE
  Transactions on Electron Devices}, vol.~60, no.~9, pp. 2808--2814, 2013.
  [Online]. Available: \url{https://doi.org/10.1109/TED.2013.2275082}
\BIBentrySTDinterwordspacing

\bibitem{manipatruni2014vector}
\BIBentryALTinterwordspacing
S.~Manipatruni, D.~E. Nikonov, and I.~A. Young, ``Vector spin modeling for
  magnetic tunnel junctions with voltage dependent effects,'' \emph{Journal of
  Applied Physics}, vol. 115, no.~17, p. 17B754, 2014. [Online]. Available:
  \url{https://doi.org/10.1063/1.4868495}
\BIBentrySTDinterwordspacing

\bibitem{jabeur2014comparison}
\BIBentryALTinterwordspacing
K.~Jabeur, F.~Bernard-Granger, G.~Di~Pendina, G.~Prenat, and B.~Dieny,
  ``Comparison of verilog-a compact modelling strategies for spintronic
  devices,'' \emph{Electronics letters}, vol.~50, no.~19, pp. 1353--1355, 2014.
  [Online]. Available: \url{https:/doi.org/10.1049/el.2014.1083}
\BIBentrySTDinterwordspacing

\bibitem{zhang2015compact}
\BIBentryALTinterwordspacing
Y.~Zhang, B.~Yan, W.~Kang, Y.~Cheng, J.-O. Klein, Y.~Zhang, Y.~Chen, and
  W.~Zhao, ``Compact model of subvolume mtj and its design application at
  nanoscale technology nodes,'' \emph{IEEE Transactions on Electron Devices},
  vol.~62, no.~6, pp. 2048--2055, 2015. [Online]. Available:
  \url{https://doi.org/10.1109/TED.2015.2414721}
\BIBentrySTDinterwordspacing

\bibitem{chen2015comprehensive}
\BIBentryALTinterwordspacing
T.~Chen, A.~Eklund, E.~Iacocca, S.~Rodriguez, B.~G. Malm, J.~{\AA}kerman, and
  A.~Rusu, ``Comprehensive and macrospin-based magnetic tunnel junction spin
  torque oscillator model-part i: analytical model of the mtj sto,'' \emph{IEEE
  Transactions on Electron Devices}, vol.~62, no.~3, pp. 1037--1044, 2015.
  [Online]. Available: \url{https://doi.org/10.1109/TED.2015.2390676}
\BIBentrySTDinterwordspacing

\bibitem{kim2015technology}
\BIBentryALTinterwordspacing
J.~Kim, A.~Chen, B.~Behin-Aein, S.~Kumar, J.-P. Wang, and C.~H. Kim, ``A
  technology-agnostic mtj spice model with user-defined dimensions for stt-mram
  scalability studies,'' in \emph{Custom Integrated Circuits Conference (CICC),
  2015 IEEE}.\hskip 1em plus 0.5em minus 0.4em\relax IEEE, 2015, pp. 1--4.
  [Online]. Available: \url{https://doi.org/10.1109/CICC.2015.7338407}
\BIBentrySTDinterwordspacing

\bibitem{camsari2015modular}
\BIBentryALTinterwordspacing
K.~Y. Camsari, S.~Ganguly, and S.~Datta, ``Modular approach to spintronics,''
  \emph{Scientific reports}, vol.~5, p. 10571, 2015. [Online]. Available:
  \url{https://doi.org/10.1038/srep10571}
\BIBentrySTDinterwordspacing

\bibitem{kazemi2016compact}
\BIBentryALTinterwordspacing
M.~Kazemi, G.~E. Rowlands, E.~Ipek, R.~A. Buhrman, and E.~G. Friedman,
  ``Compact model for spin--orbit magnetic tunnel junctions,'' \emph{IEEE
  Transactions on Electron Devices}, vol.~63, no.~2, pp. 848--855, 2016.
  [Online]. Available: \url{https://doi.org/10.1109/TED.2015.2510543}
\BIBentrySTDinterwordspacing

\bibitem{camsari2016modular}
\BIBentryALTinterwordspacing
K.~Y. Camsari, S.~Ganguly, and D.~Datta, ``A modular spin-circuit model for
  magnetic tunnel junction devices,'' in \emph{Device Research Conference
  (DRC), 2016 74th Annual}.\hskip 1em plus 0.5em minus 0.4em\relax IEEE, 2016,
  pp. 1--2. [Online]. Available: \url{https://doi.org/10.1109/DRC.2016.7673632}
\BIBentrySTDinterwordspacing

\bibitem{de2017compact}
\BIBentryALTinterwordspacing
R.~De~Rose, M.~Lanuzza, M.~d'Aquino, G.~Carangelo, G.~Finocchio, F.~Crupi, and
  M.~Carpentieri, ``A compact model with spin-polarization asymmetry for
  nanoscaled perpendicular mtjs,'' \emph{IEEE Transactions on Electron
  Devices}, vol.~64, no.~10, pp. 4346--4353, 2017. [Online]. Available:
  \url{https://doi.org/10.1109/TED.2017.2734967}
\BIBentrySTDinterwordspacing

\bibitem{modnano}
\BIBentryALTinterwordspacing
 \textit{Modular Approach to Spintronics.} Accessed: Apr. 2018. [Online]. Available: \url{https://nanohub.org/groups/spintronics}
\BIBentrySTDinterwordspacing

\bibitem{larson2000modified}
\BIBentryALTinterwordspacing
J.~D. Larson, P.~D. Bradley, S.~Wartenberg, and R.~C. Ruby, ``Modified
  butterworth-van dyke circuit for fbar resonators and automated measurement
  system,'' in \emph{Ultrasonics Symposium, 2000 IEEE}, vol.~1.\hskip 1em plus
  0.5em minus 0.4em\relax IEEE, 2000, pp. 863--868. [Online]. Available:
  \url{https://doi.org/10.1109/ULTSYM.2000.922679}
\BIBentrySTDinterwordspacing

\bibitem{cao2002predictive}
\BIBentryALTinterwordspacing
Y.~Cao, T.~Sato, D.~Sylvester, M.~Orshansky, and C.~Hu, ``Predictive technology
  model,'' \emph{Internet: http://ptm. asu. edu}, 2002. [Online]. Available:
  \url{http://ptm.asu.edu}
\BIBentrySTDinterwordspacing

\bibitem{butler2012switching}
\BIBentryALTinterwordspacing
W.~H. Butler, T.~Mewes, C.~K. Mewes, P.~Visscher, W.~H. Rippard, S.~E. Russek,
  and R.~Heindl, ``Switching distributions for perpendicular spin-torque
  devices within the macrospin approximation,'' \emph{IEEE Transactions on
  Magnetics}, vol.~48, no.~12, pp. 4684--4700, 2012. [Online]. Available:
  \url{https://doi.org/10.1109/TMAG.2012.2209122}
\BIBentrySTDinterwordspacing

\bibitem{wang2016multiphysics}
\BIBentryALTinterwordspacing
T.~Wang and J.~Roychowdhury, ``Multiphysics modelling and simulation in
  berkeley mapp,'' in \emph{Numerical Electromagnetic and Multiphysics Modeling
  and Optimization (NEMO), 2016 IEEE MTT-S International Conference on}.\hskip
  1em plus 0.5em minus 0.4em\relax IEEE, 2016, pp. 1--3. [Online]. Available:
  \url{https://doi.org/10.1109/NEMO.2016.7561645}
\BIBentrySTDinterwordspacing

\bibitem{xie2017fokker}
\BIBentryALTinterwordspacing
Y.~Xie, B.~Behin-Aein, and A.~W. Ghosh, ``Fokker---planck study of parameter
  dependence on write error slope in spin-torque switching,'' \emph{IEEE
  Transactions on Electron Devices}, vol.~64, no.~1, pp. 319--324, 2017.
  [Online]. Available: \url{https://doi.org/10.1109/TED.2016.2632438}
\BIBentrySTDinterwordspacing

\bibitem{garcia1998langevin}
\BIBentryALTinterwordspacing
J.~L. Garc{\'\i}a-Palacios and F.~J. L{\'a}zaro, ``Langevin-dynamics study of
  the dynamical properties of small magnetic particles,'' \emph{Physical Review
  B}, vol.~58, no.~22, p. 14937, 1998. [Online]. Available:
  \url{https://doi.org/10.1103/PhysRevB.58.14937}
\BIBentrySTDinterwordspacing

\bibitem{roy2016compact}
\BIBentryALTinterwordspacing
A.~S. Roy, A.~Sarkar, and S.~P. Mudanai, ``Compact modeling of magnetic
  tunneling junctions,'' \emph{IEEE Transactions on Electron Devices}, vol.~63,
  no.~2, pp. 652--658, 2016. [Online]. Available:
  \url{https://doi.org/10.1109/NEWCAS.2008.4606363}
\BIBentrySTDinterwordspacing

\bibitem{ament2016solving}
\BIBentryALTinterwordspacing
S.~Ament, N.~Rangarajan, A.~Parthasarathy, and S.~Rakheja, ``Solving the
  stochastic landau-lifshitz-gilbert-slonczewski equation for monodomain
  nanomagnets: A survey and analysis of numerical techniques,'' \emph{arXiv
  preprint arXiv:1607.04596}, 2016. [Online]. Available:
  \url{https://arxiv.org/abs/1607.04596}
\BIBentrySTDinterwordspacing

\bibitem{hong2016spin}
\BIBentryALTinterwordspacing
S.~Hong, S.~Sayed, and S.~Datta, ``Spin circuit representation for the spin
  hall effect,'' \emph{IEEE Transactions on Nanotechnology}, vol.~15, no.~2,
  pp. 225--236, 2016. [Online]. Available:
  \url{https://doi.org/10.1109/TNANO.2016.2514410}
\BIBentrySTDinterwordspacing

\bibitem{brataas2006non}
\BIBentryALTinterwordspacing
A.~Brataas, G.~E. Bauer, and P.~J. Kelly, ``Non-collinear magnetoelectronics,''
  \emph{Physics Reports}, vol. 427, no.~4, pp. 157--255, 2006. [Online].
  Available: \url{https://doi.org/10.1016/j.physrep.2006.01.001}
\BIBentrySTDinterwordspacing

\bibitem{srinivasan2014modeling}
\BIBentryALTinterwordspacing
S.~Srinivasan, V.~Diep, B.~Behin-Aein, A.~Sarkar, and S.~Datta, ``Modeling
  multi-magnet networks interacting via spin currents,'' \emph{Handbook of
  Spintronics}. Springer, pp. 1--49, 2014. [Online]. Available:
  \url{https://doi.org/10.1007/978-94-007-6892-5\_46}
\BIBentrySTDinterwordspacing

\bibitem{shi2018fast}
\BIBentryALTinterwordspacing
S.~Shi, Y.~Ou, S.~Aradhya, D.~Ralph, and R.~Buhrman, ``Fast low-current
  spin-orbit-torque switching of magnetic tunnel junctions through atomic
  modifications of the free-layer interfaces,'' \emph{Physical Review Applied},
  vol.~9, no.~1, p. 011002, 2018. [Online]. Available:
  \url{https://doi.org/10.1103/PhysRevApplied.9.011002}
\BIBentrySTDinterwordspacing

\bibitem{datta2012voltage}
\BIBentryALTinterwordspacing
D.~Datta, B.~Behin-Aein, S.~Datta, and S.~Salahuddin, ``Voltage asymmetry of
  spin-transfer torques,'' \emph{IEEE Transactions on Nanotechnology}, vol.~11,
  no.~2, pp. 261--272, 2012. [Online]. Available:
  \url{https://doi.org/10.1109/TNANO.2011.2163147}
\BIBentrySTDinterwordspacing

\bibitem{datta2017quantitative}
\BIBentryALTinterwordspacing
D.~Datta, H.~Dixit, S.~Agarwal, A.~Dasgupta, M.~Tran, D.~Houssameddine, Y.~S.
  Chauhan, D.~Shum, and F.~Benistant, ``Quantitative model for switching
  asymmetry in perpendicular mtj: A material-device-circuit co-design,'' in
  \emph{Electron Devices Meeting (IEDM), 2017 IEEE International}.\hskip 1em
  plus 0.5em minus 0.4em\relax IEEE, 2017, pp. 31--5. [Online]. Available:
  \url{https://doi.org/10.1109/IEDM.2017.8268482}
\BIBentrySTDinterwordspacing

\bibitem{double_int_2013}
\BIBentryALTinterwordspacing
H.~Sato, T.~Yamamoto, M.~Yamanouchi, S.~Ikeda, S.~Fukami, K.~Kinoshita,
  F.~Matsukura, N.~Kasai, and H.~Ohno, ``Comprehensive study of cofeb-mgo
  magnetic tunnel junction characteristics with single- and double-interface
  scaling down to 1x nm,'' in \emph{2013 IEEE International Electron Devices
  Meeting}, Dec 2013, pp. 3.2.1--3.2.4. [Online]. Available:
  \url{https://doi.org/10.1109/IEDM.2013.6724550}
\BIBentrySTDinterwordspacing

\bibitem{camsari2014physics}
\BIBentryALTinterwordspacing
K.~Y. Camsari, S.~Ganguly, D.~Datta, and S.~Datta, ``Physics-based
  factorization of magnetic tunnel junctions for modeling and circuit
  simulation,'' in \emph{Electron Devices Meeting (IEDM), 2014 IEEE
  International}.\hskip 1em plus 0.5em minus 0.4em\relax IEEE, 2014, pp. 35--6.
  [Online]. Available: \url{https://doi.org/10.1109/IEDM.2014.7047177}
\BIBentrySTDinterwordspacing

\bibitem{lin200945nm}
\BIBentryALTinterwordspacing
C.~Lin, S.~Kang, Y.~Wang, K.~Lee, X.~Zhu, W.~Chen, X.~Li, W.~Hsu, Y.~Kao,
  M.~Liu \emph{et~al.}, ``45nm low power cmos logic compatible embedded stt
  mram utilizing a reverse-connection 1t/1mtj cell,'' in \emph{Electron Devices
  Meeting (IEDM), 2009 IEEE International}.\hskip 1em plus 0.5em minus
  0.4em\relax IEEE, 2009, pp. 1--4. [Online]. Available:
  \url{https://doi.org/10.1109/IEDM.2009.5424368}
\BIBentrySTDinterwordspacing

\bibitem{kanai2012electric}
\BIBentryALTinterwordspacing
S.~Kanai, M.~Yamanouchi, S.~Ikeda, Y.~Nakatani, F.~Matsukura, and H.~Ohno,
  ``Electric field-induced magnetization reversal in a perpendicular-anisotropy
  cofeb-mgo magnetic tunnel junction,'' \emph{Applied Physics Letters}, vol.
  101, no.~12, p. 122403, 2012. [Online]. Available:
  \url{https://doi.org/10.1063/1.4753816}
\BIBentrySTDinterwordspacing

\bibitem{amiri2012voltage}
\BIBentryALTinterwordspacing
P.~K. Amiri and K.~L. Wang, ``Voltage-controlled magnetic anisotropy in
  spintronic devices,'' in \emph{Spin}, vol.~2, no.~03.\hskip 1em plus 0.5em
  minus 0.4em\relax World Scientific, 2012, p. 1240002. [Online]. Available:
  \url{https://doi.org/10.1142/S2010324712400024}
\BIBentrySTDinterwordspacing

\bibitem{liu2012spin}
\BIBentryALTinterwordspacing
L.~Liu, C.-F. Pai, Y.~Li, H.~Tseng, D.~Ralph, and R.~Buhrman, ``Spin-torque
  switching with the giant spin hall effect of tantalum,'' \emph{Science}, vol.
  336, no. 6081, pp. 555--558, 2012. [Online]. Available:
  \url{https://doi.org/10.1126/science.1218197}
\BIBentrySTDinterwordspacing

\bibitem{chen2016spin}
\BIBentryALTinterwordspacing
T.~Chen, R.~K. Dumas, A.~Eklund, P.~K. Muduli, A.~Houshang, A.~A. Awad,
  P.~D{\"u}rrenfeld, B.~G. Malm, A.~Rusu, and J.~{\AA}kerman, ``Spin-torque and
  spin-hall nano-oscillators,'' \emph{Proceedings of the IEEE}, vol. 104,
  no.~10, pp. 1919--1945, 2016. [Online]. Available:
  \url{https://doi.org/10.1109/JPROC.2016.2554518}
\BIBentrySTDinterwordspacing

\bibitem{quinsat2010amplitude}
\BIBentryALTinterwordspacing
M.~Quinsat, D.~Gusakova, J.~Sierra, J.~Michel, D.~Houssameddine, B.~Delaet,
  M.-C. Cyrille, U.~Ebels, B.~Dieny, L.~Buda-Prejbeanu \emph{et~al.},
  ``Amplitude and phase noise of magnetic tunnel junction oscillators,''
  \emph{Applied Physics Letters}, vol.~97, no.~18, p. 182507, 2010. [Online].
  Available: \url{https://doi.org/10.1063/1.3506901}
\BIBentrySTDinterwordspacing

\bibitem{torunbalci2018magneto}
\BIBentryALTinterwordspacing
M.~M. Torunbalci, T.~A. Gosavi, K.~Y. Camsari, and S.~A. Bhave, ``Magneto
  acoustic spin hall oscillators,'' \emph{Scientific reports}, vol.~8, no.~1,
  p. 1119, 2018. [Online]. Available:
  \url{https://doi.org/10.1038/s41598-018-19443-6}
\BIBentrySTDinterwordspacing

\bibitem{sayed2016multi}
\BIBentryALTinterwordspacing
S.~Sayed, S.~Hong, and S.~Datta, ``Multi-terminal spin valve on channels with
  spin-momentum locking,'' \emph{Scientific reports}, vol.~6, p. 35658, 2016.
  [Online]. Available: \url{https://doi.org/10.1038/srep35658}
\BIBentrySTDinterwordspacing

\bibitem{sayed2017transmission}
\BIBentryALTinterwordspacing
------, ``Transmission line model for charge and spin transport in channels
  with spin-momentum locking,'' \emph{arXiv preprint arXiv:1707.04051}, 2017.
  [Online]. Available: \url{https://arxiv.org/abs/1707.04051}
\BIBentrySTDinterwordspacing

\bibitem{sayed2017proposal}
\BIBentryALTinterwordspacing
S.~Sayed, S.~Hong, E.~E. Marinero, and S.~Datta, ``Proposal of a single
  nano-magnet memory device,'' \emph{IEEE Electron Device Letters}, vol.~38,
  no.~12, pp. 1665--1668, 2017. [Online]. Available:
  \url{https://doi.org/10.1109/LED.2017.2761318}
\BIBentrySTDinterwordspacing

\bibitem{lee2018multi}
\BIBentryALTinterwordspacing
J.-h. Lee, H.-j. Kim, J.~Chang, S.~H. Han, H.~C. Koo, S.~Sayed, S.~Hong, and
  S.~Datta, ``Multi-terminal spin valve in a strong rashba channel exhibiting
  three resistance states,'' \emph{Scientific reports}, vol.~8, no.~1, p. 3397,
  2018. [Online]. Available: \url{https://doi.org/10.1038/s41598-018-21760-9}
\BIBentrySTDinterwordspacing

\bibitem{vodenicarevic2017low}
\BIBentryALTinterwordspacing
D.~Vodenicarevic, N.~Locatelli, A.~Mizrahi, J.~S. Friedman, A.~F. Vincent,
  M.~Romera, A.~Fukushima, K.~Yakushiji, H.~Kubota, S.~Yuasa \emph{et~al.},
  ``Low-energy truly random number generation with superparamagnetic tunnel
  junctions for unconventional computing,'' \emph{Physical Review Applied},
  vol.~8, no.~5, p. 054045, 2017. [Online]. Available:
  \url{https://doi.org/10.1103/PhysRevApplied.8.054045}
\BIBentrySTDinterwordspacing

\bibitem{camsari2017implementing}
\BIBentryALTinterwordspacing
K.~Y. Camsari, S.~Salahuddin, and S.~Datta, ``Implementing p-bits with embedded
  mtj,'' \emph{IEEE Electron Device Letters}, vol.~38, no.~12, pp. 1767--1770,
  2017. [Online]. Available: \url{https://doi.org/10.1109/LED.2017.2768321}
\BIBentrySTDinterwordspacing

\bibitem{sutton2017intrinsic}
\BIBentryALTinterwordspacing
B.~Sutton, K.~Y. Camsari, B.~Behin-Aein, and S.~Datta, ``Intrinsic optimization
  using stochastic nanomagnets,'' \emph{Scientific Reports}, vol.~7, p. 44370,
  2017. [Online]. Available: \url{https://doi.org/10.1038/srep44370}
\BIBentrySTDinterwordspacing

\bibitem{behin2016building}
\BIBentryALTinterwordspacing
B.~Behin-Aein, V.~Diep, and S.~Datta, ``A building block for hardware belief
  networks,'' \emph{Scientific reports}, vol.~6, p. 29893, 2016. [Online].
  Available: \url{https://doi.org/10.1038/srep29893}
\BIBentrySTDinterwordspacing

\bibitem{camsari2017stochastic}
\BIBentryALTinterwordspacing
K.~Y. Camsari, R.~Faria, B.~M. Sutton, and S.~Datta, ``Stochastic p-bits for
  invertible logic,'' \emph{Physical Review X}, vol.~7, no.~3, p. 031014, 2017.
  [Online]. Available: \url{https://doi.org/10.1103/PhysRevX.7.031014}
\BIBentrySTDinterwordspacing

\end{thebibliography}


\end{document}